\documentclass[onecollarge]{svjour2}
\smartqed  
\usepackage{graphicx}
\usepackage{mathptmx}
\usepackage{amssymb}
\usepackage{amsmath}
\usepackage{graphicx}
\usepackage{bm}
\journalname{Few-Body Systems}
\begin{document}

\title{Weakly bound LiHe$_2$  molecules in the framework of  three-dimensional Faddeev equations}

\titlerunning{Weakly bound Li He$_2$ molecules ...}     

\author{E. A. Kolganova \and V. Roudnev }

\authorrunning{E.\,A.\,Kolganova, V.\,Roudnev} 

\institute{E. A. Kolganova   \at
              Bogoliubov Laboratory of Theoretical Physics, Joint  Institute for Nuclear Research  
							and  Dubna State University, 141980 Dubna, Russia
							\email{kea@theor.jinr.ru}    
							\and
							V. Roudnev \at
							Department of Computational Physics, St Petersburg State University, 7/9
              Universitetskaya nab., St Petersburg, 199034, Russia
          }


\maketitle

\begin{abstract}
A method of direct solution of the Faddeev equations for the bound-state 
problem  with zero total angular momentum is used to
calculate the binding energies. The results for binding energies of He$_2$$^6$Li 
and He$_2$$^7$Li systems and helium atom - HeLi dimer scattering length are 
presented. The results show that modern potential models support two bound states in both 
trimers. In both cases the energy of the excited state is very close to the 
energy of the lowest two-body threshold. 
\keywords{Efimov effect \and triatomic systems \and Faddeev approach \and helium-alkali trimers}
\end{abstract}

\bigskip

\section{Introduction}

The Efimov effect is a remarkable phenomenon, which is an excellent illustration 
of the variety of possibilities
arising when we transit from the two-body to the three-body problem. In 1970 
V.Efimov~\cite{Efimov1970_PLB,Efimov1970_YAF} proposed that three-body systems 
with short range interaction can have an infinite number of bound states when 
none of the two-particle subsystems has bound states but at least two of them 
have infinite scattering lengths. In such a case the scattering length $a$ is 
much lager than the range of the interaction $r_0$. The simplest situation described 
by Efimov~\cite{Efimov1970_YAF} corresponds to three identical neutral bosons 
interacting via short-range resonant interactions treated in the  zero-range 
theory framework. In this theory it is assumed that the short-range region details 
of the interaction can be neglected and the wave function in the
asymptotically free region can be parametrized by the scattering length. In 
order to reproduce correctly the two-body wave function in the region 
outside of the range of interaction $r_0$  one can use the Bethe-Peierls 
boundary conditions~\cite{BethePeierls} for the three-body wave function $\Psi$ 
when the two particles separated by $r$ come in contact
\begin{equation}
- \frac{1}{r\Psi} \frac{\partial r \Psi}{\partial r} \xrightarrow[{r \rightarrow 
0}]{} \frac{1}{a} .
\label{BethePeierls0}
\end{equation}
The simplification which is used in the zero-range theory is to keep the same 
form of the wave function down to $r=0$, although this is unphysical at 
distances $r < r_0$.
To describe the three-body system Efimov used the free Schr\"odinger equation 
written in hyperspherical coordinates~\cite{Delves} with Bethe-Peierls boundary 
conditions (\ref{BethePeierls0})  which lead to the equation for a radial function
\begin{equation}
\left( - \frac{d^2}{d R^2} - \frac{1}{R}  \frac{d}{d R} + \frac{s_n^2}{R^2} 
\right) F_n (R) = E F_n (R) ,
\label{RadEq}
\end{equation}
where $R$
is the hyperradius and $s_n$ is a solution of the following transcendental 
equation~\cite{Efimov1970_YAF,Danilov}
\begin{equation}
 - s_n \cos\left(s_n \frac{\pi}{2} \right) + \frac{8}{\sqrt{3}} \sin \left( s_n 
\frac{\pi}{6}\right) = 0.
\label{TrEq}
\end{equation}
All the solutions of equation (\ref{TrEq}) are real, except one $s_0 = 1.0062378 i$ 
which is purely imaginary, which results in an attractive effective potential in equation 
(\ref{RadEq}) for $n=0$.
This attraction is the origin of the Efimov effect. In order to prevent the Thomas 
collapse~\cite{Thomas} an additional three-body boundary 
condition can be used to fix some of the three-body observables (the ground 
state energy or the particle-dimer scattering  length). This boundary condition 
breaks the scale invariance under arbitrary scale transformations but still 
keeps the scale invariance under some discrete set of scale transformation with 
scaling factors being powers of $\lambda=\exp{(\pi/|s_0|)}$. Thus, in the limit 
$a \rightarrow \infty$ there is an infinite number of bound states, forming a 
geometric series of energies accumulated at the threshold. The following relationship holds
for three identical bosons 
\begin{equation}
\lim\limits_{n\to\infty}\dfrac{E_{n+1}}{E_n}=\exp(-{2\pi}/{|s_0|})\equiv \lambda^{-2} \approx 
\frac{1}{515.035}\approx \frac{1}{22.7^2}.
\label{Efimrel}
\end{equation}

One of the best theoretically predicted three-body system with an excited state 
of the Efimov type is a naturally existing molecule of the helium trimer 
$^4$He$_3$ (see,~\cite{FBS2011,FBS2017} and refs. therein). The interaction between 
two helium atoms is quite weak and supports only one 
bound state with the energy about 1mK and a rather large scattering length 
about 100 \AA. Only recently the long predicted weakly-bound excited state of the helium trimer 
was observed for the first time using a combination of Coulomb explosion imaging and cluster 
mass selection by matter wave diffraction~\cite{Kunitski}.

The first experimental evidence of the Efimov resonance was observed in an 
ulracold gas of $^{133}$Cs atoms in 2006~\cite{Kraemer}. Experimentally, they 
observed a giant three-body recombination loss when the strength of two-body 
interaction was varied. More recently, the second Efimov resonance has been 
observed and the scaling factor for the Efimov period  has been found to be 
$21.0$~\cite{Huang2014,Huang2015}, close to the universal ratio $\lambda = 22.7$ 
in a homonuclear system~\cite{Efimov1970_PLB}. It was shown in 
\cite{Efimov1973,Esry2006,Wang2012} that the universal Efimov scaling is also 
valid for systems of non-identical particles. In particular,  for a system composed of 
two heavy and one light atom the scaling factor $\lambda$ (it depends only on the masses of the particles) 
gets smaller as the mass imbalance increases.
Experimentally, heteronuclear Efimov states have been searched for in  
$^{41}$K$^{87}$Rb$_2$~\cite{Barontini},
 $^{40}$K$^{87}$Rb$_2$~\cite{Bloom},  $^{39}$K$^{87}$Rb$_2$~\cite{Wacker},
 $^{7}$Li$^{87}$Rb$_2$~\cite{Maier},  $^{6}$Li$^{133}$Cs$_2$~\cite{Tung,Ulmanis} 
systems.
Recent reviews on Efimov effect could be found in~\cite{Naidon,Greene}.

There is a growing interest in the investigation of He$_2$ - alkali-atom 
van-der-Waals systems, that are expected to be of Efimov nature. In addition to 
the Helium dimer, the He - alkali-atom interactions are even shallower and also 
support weakly bound states. In triatomic $^4$He$_2$-alkali-atom systems presence 
of Efimov levels can be expected. Three-body recombination and 
atom-molecular collision in Helium-Helium-alkali-metal systems
at ultracold temperatures have been studied using adiabatic hyperspherical 
representation in Ref.~\cite{Suno09,Suno17,Wu}.  Here we use the Faddeev 
equations in total angular momentum representation to calculate the 
$^4$He$_2$$^{6;7}$Li binding energies and a scattering length, which has not 
been studied  before.

\section{Method}
\label{Sdt}

The configuration space of three particles after elimination of the center of 
mass can be described in terms of three sets of Jacobi coordinates
\begin{equation}
\begin{array}{rcl}
           {\bf x}_i&=&\displaystyle\left(\frac{2m_j m_k }
                {m_j+m_k}\right)^{1/2}
                ({\bf r}_j-{\bf r}_k)\cr
        {\bf y}_i&=&\displaystyle\left(\frac{2m_i (m_j +m_k )}
             {m_i +m_j +m_k }\right )^{1/2}
                \left({\bf r}_i-\displaystyle
               \frac{m_j {\bf r}_j+m_k
                {\bf r}_k}{m_j+m_k }\right)
\end{array}
\label{Jacoord}
\end{equation}
The set of coordinates $i$ describes a partitioning of the three particles into 
a pair $(jk)$ and a separate particle $i$.  The Jacoby vectors with different indexes 
are related by an orthogonal transform
\begin{equation}
\label{eq:rotation}
 {\bf x}_j =  {\sf c}_{ji}{\bf x}_i+  {\sf s}_{ji}  {\bf y}_i \quad {\bf y}_j = 
-{\sf s}_{ji}  {\bf x}_i  + {\sf c}_{ji}{\bf y}_i  \,.
\end{equation}
The coefficients $c_{ji}$ and $ s_{ji}$ are expressed through the masses of the 
particles
\begin{eqnarray*}
 {\sf c}_{ji}&=&-\left(
\frac{{\rm m}_i {\rm m}_j}
    {({\rm m}_i + {\rm m}_k)
    ({\rm m}_j + {\rm m}_k)} \right)^{1/2}, \quad
    {\sf s}_{ji}=(-1)^{j-i}
\mathop{\rm sign}(j-i)
\left( 1-{\sf c}_{ji}^2 \right)^{1/2}
\end{eqnarray*}
and satisfy $c^2_{ji}+s^2_{ji}=1$.

The three-body system is described by the Hamiltonian
\begin{equation}
\label{eq:Hamiltonian}
        H=H_0+\sum _i V_i ({\bf x}_i )\,,
\end{equation}
where $H_0$ stands for the kinetic energy of the three particles and $V_i ({\bf x}_i )$ is
the interaction potential acting in the pair $i$. Faddeev decomposition 
represents the wave function $\Psi$ in terms of  the Faddeev components $\Phi_i$
\begin{equation}
  \Psi=\sum_i  \Phi_i({\bf x}_i, {\bf y}_i) \, ,
  \label{eq:WF}
\end{equation}
which satisfy the following set of equations~\cite{MerkFadd}
\begin{equation}
(-\Delta_{{\bf x}_i} -\Delta_{{\bf y}_i}+ V_i({\bf x}_i) - E) \Phi_i({\bf x}_i, 
{\bf y}_i) =
  V_i({\bf x}_i) \sum_{k\ne i}  \Phi_k({\bf x}_k, {\bf y}_k) \ ,
\label{eq:Faddeev}
\end{equation}
where $E$ is the total energy of the system.
In case of zero total angular momentum the angular degrees of freedom 
corresponding to collective rotation of the three-body system can be separated \cite{Kvits} and 
the kinetic energy operator reduces to
\begin{equation}
\label{H00}
        H_{0}=-\frac{\partial ^{2}}
        {\partial x_i^{2}}-\frac{\partial ^{2}}
        {\partial y_i^{2}}-(\frac{1}{x_i^{2}}
        +\frac{1}{y_i^{2}})\frac{\partial }
        {\partial z_i}(1-z_i^{2})
        \frac{\partial }{\partial z_i}\,,
\end{equation}
where $x_i$, $y_i$ and $z_i$ are intrinsic coordinates
\begin{equation}
\label{IntrCoord}
        x_i=|{\bf x}_i|,\quad y_i=|{\bf y}_i|,\quad
                z_i=\frac{({\bf x}_i,{\bf y}_i)}{x_iy_i},\quad
                x_i,y_i\in [0,\infty ),\ \  z_i\in [-1,1]  .
\end{equation}
Due to (\ref{Jacoord}) and (\ref{eq:rotation}) these coordinates are related by
\begin{eqnarray*}
x_{j}&=&\sqrt{{\sf c}_{ji}^2 x^2_i+
             2{\sf c}_{ji}{\sf s}_{ji} x_iy_i z_i
             +{\sf s}_{ji}^2 y_i^2}, \quad
y_{j}=\sqrt{{\sf s}_{ji}^2 x_i^2
          -2{\sf c}_{ji}{\sf s}_{ji} x_iy_iz_i
          +{\sf c}_{ji}^2 y_i^2},\\
& & x_j y_j z_j = ({\sf c}_{ji}^2 - {\sf s}_{ji}^2)x_iy_i z_i - 
                   {\sf c}_{ji}{\sf s}_{ji} (x_i^2 - y_i^2)\,.
 \end{eqnarray*}
As a result we have a set of three-dimensional differential Faddeev equations
\begin{equation}
\label{EqFaddTAM1} \displaystyle
   (H_0+V_i(x_i)-E)\phi_i(x_i,y_i,z_i)
   =  -V_i(x_i) \sum_{k\ne i} \phi_k(x_k,y_k,z_k) \,.
\end{equation}
When two particles of a system are identical, the Faddeev equations can be 
simplified.
For example, for the  He$_2$Li atomic systems particles 1 and 2 
corresponding to $^4$He atoms are
identical and the Faddeev components $\phi_1(x_1,y_1,z_1)$ and $\phi_2(x_2,y_2,z_2)$ 
 transform
into each other under an appropriate rotation of the coordinate space. 
Therefore, it is sufficient to consider
only two independent Faddeev components.
In the case of three identical bosons all the Faddeev components take identical 
functional form, which
makes it possible to reduce the system of three equations (\ref{EqFaddTAM1}) to 
one equation.

Using the fact that the both dimers -- $^4$He$_2$ and $^4$HeLi --  have a unique bound state, 
the asymptotic boundary condition for a bound state as 
$\rho=\sqrt{x^2+y^2}\rightarrow\infty$ and/or $y\rightarrow\infty$
reads as follows (see \cite{MerkFadd,kea17})
\begin{equation}
\label{HeBS}
        \begin{array}{rcl}
 \phi(x,y,z)  =   \psi_d(x)\exp({\rm i} \sqrt{E-\epsilon_d}\,y) {\rm a}_0 (z)
        + \displaystyle\frac{\exp({\rm i}\sqrt{E}\rho)}{\sqrt{\rho}} A(y/x,z),
\end{array}
\end{equation}
where $\epsilon_d$ stands for the corresponding dimer energy while $\psi_d(x)$ 
denotes the dimer
wave function.  The coefficients ${\rm a}_0$ and $A(y/x,z)$ describe 
contributions into
$\phi(x,y,z)$ from (2+1) and (1+1+1) channels respectively. The last term can 
be neglected for the states below the three-body threshold. 
For bound state calculations Dirichlet or Neumann boundary conditions can also 
be employed.

\section{Results and Discussions}
\label{Salk}

The Li-He interaction is described by the  KTTY potential~\cite{KTTY}, 
theoretically derived  by  Kleinekath\"ofer, Tang, Toennies and Yiu with more 
accurate coefficients taken from \cite{Yan,KLM-KTTY}.
Calculated values of the binding energy for
$^4$He$^6$Li is 1.512 mK and for $^4$He$^7$Li is 5.622 mK.  Such small values of 
binding energy give indication on possible existence of Efimov states in the
corresponding He$_2$Li triatomic systems.

In our calculations we use two different model potentials for He-He interaction - 
TTY~\cite{TTY} and Przybytek~\cite{PRZ2010}. The purely theoretical TTY 
potential derived by Tang, Toennies and Yiu~\cite{TTY} is  based  on  the  
perturbation  theory and is
described by a relatively simple analytical expression. The recent 
Przybytek~\cite{PRZ2010} potential includes relativistic and quantum 
electrodynamics contributions as well as some accuracy improvements.
Each of these potentials supports a single weakly bound state of the dimer.
Calculated values of the binding energy $\epsilon_d$ for the corresponding 
dimers, the inverse wavenumber $\kappa^{-1}= (\hbar^2/2\mu \epsilon_d)^{1/2}$  
($\mu$ is a reduced mass) and the atom-atom  scattering length $a^{2}$ are 
presented in Table~\ref{T2body}. Atomic masses for different isotopes are taken 
from~\cite{Mills}.

\begin{table}[htb]
\caption{\label{T2body}Absolute values of dimer energies $|\epsilon_d|$  (in 
mK), the inverse wavenumber $\kappa^{-1}$ ( in \AA ) and the scattering length 
$a^{2}$ (in \AA ) for He-He and $^4$He$-$alkali-atoms calculated for the 
potentials used.
}
\begin{center}
\begin{tabular}{cccc|cccc}
\hline
 Dimer & \quad $|\epsilon_d|$ (mK)\quad & $\kappa^{-1}$ (\AA ) & $a^{2}$ (\AA )&  Dimer &  \quad $|\epsilon_d|$ (mK)\quad & $\kappa^{-1}$ (\AA ) &$a^{2}$ (\AA )\\
\hline
  $^4$He$^4$He $^a$ & 1.321 & 95.78 & 99.59       				&$^4$He$^{23}$Na   & $28.97$  & 15.7 &23.37\\
$^4$He$^4$He  $^b$  & 1.620 & 86.49 & 90.28   				&$^4$He$^{39}$K    & $11.20$  & 24.4 &33.32\\
 $^4$He$^6$Li       & 1.515 & 81.63 & 89.42     				&$^4$He$^{85}$Rb   & $10.27$  & 24.9 &34.02\\
 $^4$He$^7$Li       & 5.622 & 41.14 & 48.84     				&$^4$He$^{133}$Cs  & $4.945$  & 35.5 &45.32\\
\hline
\end{tabular}
\end{center}

$^a$ Using the He-He potential from Ref.~\cite{TTY}\\
$^b$ Using the He-He potential from Ref.~\cite{PRZ2010}\\
\end{table}

From Table~\ref{T2body} we can see that the inverse wave number is a good 
approximation for the scattering length, which indicates that the zero-range potential model
(\ref{BethePeierls0}) is applicable.
The calculated binding energy of the helium dimer with Przybytek 
potential~\cite{PRZ2010} is very close to its recent experimental value
of  $-151.9 \pm 13.3$ neV $\approx -1.76 \pm 0.15$ mK~\cite{Zeller} while the 
energy obtained with TTY potential~\cite{TTY} is closer to the previous 
experimental estimation of $-1.1^{+0.3}_{-0.2}$ mK  $\approx -95^{+25}_{-15}$ 
neV~\cite{Grisenti-exp-2000}.
The choice of the He-He potential is especially important for $^4$He$^6$Li, as
switching between the two model potentials swaps the order of the two-body thresholds in the system. 
In the case of TTY potential the lowest two-body threshold corresponds to the 
$^6$LiHe system, while for the  Przybytek potential it is the 
$^4$He$_2$ dimer which is bound stronger.

All He-alkali-atom dimers are weakly bound, but the binding energies of HeLi and 
HeCs systems are of the same order as the binding energy of the Helium dimer.
It suggests that in the corresponding He$_2$-alkali-atom triatomic systems 
Efimov states might exist.
Indeed, in calculations~\cite{Suno09,Suno17,Wu,kea17} an excited state located 
very close to the HeLi threshold has been found.

To calculate the binding energy of $^4$He$_2$$^{6}$Li  and $^4$He$_2$$^{7}$Li  
trimers, we employed the equations (\ref{EqFaddTAM1}), and the bound-state 
asymptotic boundary condition (\ref{HeBS}).
The details of the numerical procedure are described 
in~\cite{Sofianos,Roudnev,PAN2012}. The three-body interaction is expected to be 
small as in the case of helium trimer~\cite{Cencek} and we do not take it into 
account.
Convergence tables for bound states of the $^4$He$_2$$^7$Li trimer
calculated with the Przybytek potential~\cite{PRZ2010} are shown in 
Table~\ref{Conv1}. The bound state energies (with respect to the three-body break-up threshold) are 
presented for different numbers of grid points. The number
of grid points in coordinates $x$ and $y$ are set equal, and the number of 
grid points in angular coordinate $z$ is varied independently. As it is seen from 
Table~\ref{Conv1}, the excited state is much less sensitive to the angular grid. 
Similar behavior had been observed for the helium  trimer~\cite{Sofianos,Roudnev}.

\begin{table}[htb]
\caption{\label{Conv1} Convergence for the $^4$He$_2$$^7$Li bound states
energies with respect to the number of grid points for the Przybytek 
potential~\cite{PRZ2010}.
}

\begin{center}
\begin{tabular}{ccccccc}
\hline
\multicolumn{7}{c} {}   Ground state energy (mK)  \\
\hline
$N_x$=$N_y$ & $N_z = 1$ & $N_z = 2$ & $N_z = 3$ & $N_z = 4$ &$N_z = 5$ & \\
\hline
 20 & -78.4124 &  -80.0524 & -80.4523 & -80.6422 & -80.6622 & \\
30 &  -78.3750 & -79.9701 & -80.3877 & -80.5630 & -80.5937 & \\
40 & -78.3849 & -79.9695 & -80.3853 & -80.5690 & -80.5977 & \\
50 & -78.3877 & -79.9727 & -80.3869 & -80.5712 & -80.5990 & \\
\hline
\multicolumn{7}{c} {}    Excited state energy (mK)  \\
\hline
 20 & -5.6457 &  -5.6522 & -5.6538 & -5.6545 & -5.6546 & \\
30 &  -5.6447 & -5.6510 & -5.6527 & -5.6534 & -5.6535 & \\
40 & -5.6448 & -5.6511 & -5.6527 & -5.6534 & -5.6536 & \\
50 & -5.6448 & -5.6511 & -5.6527 & -5.6535 & -5.6536 & \\
\hline
\end{tabular}
\end{center}
\end{table}

\begin{table}[htb]
\caption{\label{tableBS}
Bound-state energies for the He$_2$Li systems and  helium atom - HeLi dimer 
scattering length $a^3$ . The energies are given in mK and are relative to the 
three-body dissociation threshold.  The scattering length is given in \AA. The 
present results are compared to results given in references.
}
\begin{center}
{
\begin{tabular}{cc@{  }c@{  }c@{  }c@{  }c@{  }c@{  }c@{ }c@{  }c@{  }c}
\hline
 E (mK)/$a^{3}$ (\AA)      & present &present &~\cite{Suno17}&~\cite{Wu} 
&~\cite{Suno13} & ~\cite{Suno13} &  ~\cite{Suno10} &~\cite{Baccarelli} 
&~\cite{DiPaola} &~\cite{Stipanovic} \\
He-He potential& TTY &   Przybytek & LM2M2 & LM2M2 & Jeziorska & Jeziorska &  LM2M2
&  LM2M2 & TTY & HFDB\\
He-Li potental & KTTY & KTTY& KTTY & KTTY &  KTTY &  KTTY & Cvetko & Cvetko & KTTY  & KTTY\\
\hline
 & & & & & & & & &\\[-1ex]
$|$E$_{^{7}Li^{4}He_{2}}|$ & 79.35 & 80.60& 79.36 &78.73 & 76.32 &81.29 & 64.26 
& 73.3 & 80.0 & 81,03  \\
$|$E$^{*}_{^{7}Li^{4}He_{2}}|$  & 5.642  & 5.654 & 5.642 & 5.685 & 5.51 & 5.67  
& 3.01 &12.2 &   &  \\
$a^{3}$  & 683  & 553 &  &  &  &   &  &  &   &  \\[-1ex]
 & & & & & & & & &\\[-1ex]
\hline
 & & & & & & & & &\\[-1ex]
$|$E$_{^{6}Li^{4}He_{2}}|$ & 57.24 &  58.38 & 57.23 &   & & 58.88&  &   51.9 &  
& 58,72   \\
$|$E$^{*}_{^{6}Li^{4}He_{2}}|$  & 1.940 & 2.049& 1.937 &  &  &2.09 &  & 7.9 &    
&  \\
$a^{3}$  & 191  & 144$^*$  &  &  &  &   &  &  &   & 
 \\[-1ex]
 & & & & & & & & &\\[-1ex]
\hline
\end{tabular}
}
\end{center}
\texttt{}  $^*$ $^6$Li atom - He$_2$ dimer scattering length
\end{table}

Our results for $^4$He$_2$$^{7}$Li and $^4$He$_2$$^{6}$Li trimers binding 
energies as well as the results obtained by other authors are summarized in 
Table \ref{tableBS}. The results show that the both potential models support two 
bound states in the both trimers. The energy of the excited state is very close to 
the energy of the lowest two-body threshold.  Different He-He potentials give
$0.3$ mK difference in the helium dimer binding energy, which leads to the $\sim 
1.2$mK difference in the binding energy of the ground state of He$_2$Li trimers, 
although the energy of excited state of He$_2$$^7$Li is practically unchanged 
(difference is $\sim 0.01$mK). As is mentioned above, for the He$_2$$^6$Li system the lowest 
threshold is different for different potentials: for TTY it corresponds 
to the energy of HeLi bound state, while for Przybytek it is the bound state energy of the He$_2$ dimer. 
So, for different potentials the absolute value of the excited state 
energy changes slightly, but the relative energy with respect to the two-body 
threshold remains practically the same.

Results of other authors in Table \ref{tableBS} are based on solving the Shr\"odinger equation.
The adiabatic hyperspherical approach has been employed in~\cite{Suno09,Suno17,Wu,Suno13,Suno10}
and variational calculations has been performed in~\cite{Baccarelli,DiPaola,Stipanovic}. 
The fourth column of Table~\ref{tableBS} contains the results obtained by Suno
in~\cite{Suno17} using the adiabatic hyperspherical method. For the He-Li 
interaction he has used KTTY potential~\cite{KTTY} as in our calculations, but for 
the He-He interaction the LM2M2 potential~\cite{LM2M2} has been used. However, LM2M2 
and TTY potentials support $^4$He$_2$ with the energies $\varepsilon_d = -1.31$ mK and 
 $\varepsilon_d = -1.32$ mK, correspondingly \cite{Roudnev}. So good agreement between our results and results from ~\cite{Suno17} 
are not surprising (see columns 1 and 3 in Table~\ref{tableBS}).

The fifth column of Table~\ref{tableBS} contains the results obtained by Wu {\it 
et al.} \cite{Wu} using the mapping method within the adiabatic hyperspherical 
framework~\cite{Kokoouline}. The next two columns are the 
results of calculations by H. Suno,  E. Hiyama and M. Kamimura~\cite{Suno13} 
using the Gaussian expansion method and the adiabatic hyperspherical 
representation respectively, although with different He-He potentals. 
They employed the He-He potential suggested by Jeziorska \textit{et 
al.}~\cite{Jeziorska}, which gives the helium dimer biding energy $-1.74$ mK which is 
lower than for Przybytek potential. The two methods give different results, 
but authors in~\cite{Suno13} mentioned that the adiabatic 
hyperspherical representation was less accurate.
The next column is the results of calculations by Suno and 
Esry~\cite{Suno09,Suno10} by the adiabatic hyperspherical method. They also 
employed the He-He potential from~\cite{Jeziorska}, but  for 
Li-He interaction Cvetko potential from~\cite{Cvetko} has been used. 
The potential proposed by Cvetko \textit{et al.}~\cite{Cvetko} gives 
the HeLi smaller binding energy than the KTTY potential, 
namely $-2.8$ mK for $^4$He$^7$Li and $-0.31$ mK in case of $^4$He$^6$Li dimer. 
The ninth column contains the results obtained by Baccarelli 
 \textit{et al.}~\cite{Baccarelli} with the same potential as in~\cite{Suno10}, 
but using a different computation method - variational calculations in terms of 
distributed Gaussian functions.  The last two columns contain the results of 
Monte Carlo calculations by Di Paola  \textit{et al.}~\cite{DiPaola} and  
Stipanovi\'c \textit{et al.}~\cite{Stipanovic} using TTY~\cite{TTY} and HFDB~\cite{HFDB} 
as He-He interactions.

We should also mention the first results obtained by Yuan and Lin~\cite{Yuan} 
using the adiabatic hyperspherical method which gives an upper bound to the 
ground state $-45.7$ mK for $^4$He$_2$$^7$Li and $-31.4$ mK for $^4$He$_2$$^6$Li 
and the prediction of the bound state energies made by Delfino \textit{et 
al.}~\cite{Delfino} using the scaling ideas and zero-range model calculations. 
The preliminary Faddeev calculation using bipolar partial-wave expansion for 
searching Efimov states in $^4$He$_2$$^7$Li system have been performed 
in~\cite{kea17}. In these papers, however, the contribution of higher partial waves was 
underestimated because of the computational restrictions.

As it has been demonstrated in~\cite{kea17}, the excited state of He$_2$$^7$Li has a 
Efimov-type behavior similar to helium trimer system~\cite{He3-Rev}.  To check for the Efimov-like state 
the original Li-He potential has been multiplied by a factor $\lambda$. 
An increase of the coupling constant $\lambda$ makes the potential more attractive 
and Efimov levels become weaker and disappear with further increase of 
$\lambda$. Indeed this situation is observed for the excited state energy of 
He$_2$$^7$Li in contrast to the ground state energy whose absolute value 
increases continuously with increasing attraction.

The results for the He-atom -- HeLi-dimer scattering length are presented in the 
last line of the Table~\ref{tableBS} for each Li isotopes.  The helium atom -- 
helium-alkali-atom collisions at  ultralow energies are studied in 
Ref.~\cite{Suno14} by Suno and Esry using the adiabatic hyperspherical 
representation. In particular, they calculated the total cross section also for 
$^4$HeLi + $^4$He $\rightarrow$ $^4$HeLi + $^4$He elastic scattering. Our 
estimation of the cross section $\sigma=4\pi a^2$ at the threshold is $ 
5.8\times 10^{-10}$ cm$^2$ using TTY potential~\cite{TTY} and $3.8\times 
10^{-10}$ cm$^2$ using Przbytek potential~\cite{PRZ2010}. These values agree
with the value $\approx 3 \times 10^{-10}$ cm$^2$ obtained in 
Ref.~\cite{Suno14} using SAPT potential~\cite{Jeziorska} at the energy $10^{-3}$~mK 
above the threshold.

\section{Conclusion}

We have used direct solution of the Faddeev equations for the 
bound-state and scattering problems with zero total angular momentum. The numerical algorithm is 
based on spline expansion of the Faddeev components combined with the tensor trick 
preconditioning and the Arnoldi algorithm for eigenanalysis.
Calculations of the He$_2$$^6$Li and He$_2$$^7$Li ground and excited states show 
that the method is very efficient and allows one to obtain stable convergent results. 
Apparently, it performs better than the previously exploited method of the 
bipolar partial-wave expansion. 
Our results for $^4$He$_2$$^{7}$Li and $^4$He$_2$$^{6}$Li trimers binding 
energies show that different potential models support two bound states in both 
trimers. The energy of the excited state is very close to the energy of the 
lowest two-body threshold. In case of the He$_2$$^6$Li system the lowest 
threshold is different for different potentials but the relative energy with 
respect to the lowest two-body threshold is practically the same.

\begin{acknowledgements}
One of the author (EAK) would like to thank T. Frederico, A.~Kievsky and 
P.~Stipanovi\'c for stimulating discussions and also W.~Sandhas and 
A.K.~Motovilov  for their constant interest to this work.
\end{acknowledgements}

\end{document}